\documentclass[cameraready]{Interspeech}

\title{From Text Metrics to Model Internals: A Study of Whisper ASR Hallucination Detection}

\author[orcid=0000-0002-3897-181X, correspondingauthor]{Jan}{Jasiński}
\author[orcid=0009-0004-3021-7814]{Mateusz}{Barański}
\author[orcid=0000-0002-6708-876X]{Julitta}{Bartolewska}
\author[orcid=0000-0003-2921-6502]{Marcin}{Witkowski}
\author[orcid=0000-0002-7834-6920]{\\Konrad}{Kowalczyk}

\address{Signal Processing Group, Institute of Electronics, AGH University of Krakow, Poland}

\email{jjasinsk@agh.edu.pl, mbaranski@agh.edu.pl, bartolew@agh.edu.pl, witkow@agh.edu.pl, konrad.kowalczyk@agh.edu.pl}


\usepackage{comment}

\begin{document}
\maketitle

\begin{abstract}

    Hallucinations of ASR models — fluent transcriptions with no basis in audio — degrade system performance and pose risks in downstream applications. Robust detection of such errors remains a challenge. This paper studies Whisper large v3 hallucination detection on real-speech human-annotated data across three paradigms: text-based, LLM-based, and internal decoder state probing. Text classifiers utilizing metrics for text evaluation achieve high recall but degrade without reference transcripts. LLM-based detection improves precision with domain-specific prompt conditioning, yet remains less competitive than the lightweight text-based methods. Probing Whisper's decoder representations, without a ground-truth reference, yields the strongest performance, revealing that hallucination traits are encoded across intermediate decoding layers. A late-fusion meta-classifier combining text and internal-state outputs achieves the best overall detection performance.
    
\end{abstract}

\section{Introduction}

Hallucinations in Automatic Speech Recognition (ASR) systems present a critical challenge, amplified by the widespread deployment of large-scale models trained on weakly supervised data \cite{radford2023robust, Granite}. As these training datasets often contain weak or machine-generated labels, models can learn incorrect acoustic event matchings, producing fluent text that has no phonetic connection to the spoken input \cite{baranski2025investigationwhisperasrhallucinations}. Beyond degrading performance metrics like Word Error Rate (WER) \cite{frieske2024hallucinationsneuralautomaticspeech}, hallucinations are dangerous due to their unpredictable and highly plausible nature. They can inject offensive, stereotypical, or factually incorrect content into transcriptions \cite{baranski2025investigationwhisperasrhallucinations, Koenecke_2024}. When a hallucinating ASR model serves as the frontend for downstream Natural Language Processing (NLP) applications, these coherent but fabricated errors can trigger cascading system failures \cite{szymanski-etal-2023-arent}.

Currently, efforts to quantify ASR hallucinations rely heavily on traditional error metrics such as WER, Character Error Rate \cite{fang2025fewerhallucinationsverificationthreestage}, Insertion Error Rate \cite{whisperx}, alongside semantic distances like SeMaScore \cite{Sasindran_2024} and BERTScore \cite{bert}. While these metrics can be valuable as general proxies for model performance, recent benchmarks have demonstrated that they alone are insufficient to distinguish hallucinations from other errors \cite{koudounas2025hallucinationbenchmarkspeechfoundation}. Furthermore, the lack of human-annotated real-speech data has limited their validation to non-speech or synthetic audio, leaving the capacity of these metrics to distinguish hallucinations from phonetic mishearings in spontaneous speech largely unverified \cite{baranski2025investigationwhisperasrhallucinations, calmwhisper}. Finally, these metrics are defined in an oracle setting, requiring a ground-truth reference transcript, confining their use to model evaluation and development. This is reflected in SHALLOW, the recent state-of-the-art evaluation framework \cite{koudounas2025hallucinationbenchmarkspeechfoundation}, which provides a rigorous methodology for the classification of hallucination errors without directly addressing detection accuracy and usage in deployment. On the other hand, reference-free text metrics, e.g., generation speed \cite{cps}, 5-gram repetition rate \cite{whisperx}, perplexity \cite{perplexity}, struggle to match oracle fidelity. Recently, Large Language Models (LLMs) have demonstrated promise in zero-shot ASR error correction \cite{fang2025fewerhallucinationsverificationthreestage, pulikodan2025approachmeasuringperformanceautomatic}. Their application to hallucination detection has relied on prompting with both prediction and ground-truth \cite{atwany_lost_2025} confining them to the oracle setting. In parallel, emerging research on ASR interpretability shows that the internal representations of an encoder-decoder model inherently encode transcription confidence, enabling the distinction between low and high WER predictions without text references \cite{atwany_lost_2025, glazer2025transcriptionmechanisticinterpretabilityasr}. 
Expanding this concept to isolate the specific hallucination error mode could open a promising avenue for fast, reference-free ASR hallucination detection. %

In this study, we comprehensively investigate ASR hallucination detection across these three paradigms, i.e., text-based, LLM-based and decoder state-based, introducing targeted improvements to each and demonstrating how detection efficiency can be optimized. We focus on the Whisper large v3 model \cite{radford2023robust} and utilize HALAS \cite{baranski2026halas} an ASR hallucination dataset, which contains human annotations for real-speech model predictions. Our main contributions are as follows:
\begin{itemize}
\item \textbf{Text-Based Detection:} We present a systematic evaluation of oracle and reference-free textual features for hallucination detection.
We show that tree-based ensemble classifiers substantially outperform linear baselines, and that reference-free text features alone remain insufficient for reliable detection.
\item \textbf{LLM-Based Detection:} We demonstrate that out-of-the-box LLMs exhibit limited recall despite strong general language understanding, and that injecting model-specific pathology knowledge into the prompt substantially improves precision. However, LLM-based detection remains less competitive than lightweight text-based classifiers.
\item \textbf{Internal State Probing:} We show that hallucination signals are encoded across intermediate layers of the Whisper decoder and that probing full decoding sequences consistently outperforms single-token embeddings generated at the end of the decoding process, yielding the strongest reference-free performance in our study.
\item \textbf{Cross-Paradigm Analysis and Fusion:} We show that text-based and internal-state detectors capture partially non-overlapping hallucination signals, and that
a meta-classifier leveraging this complementarity achieves the best overall performance.
\end{itemize}

\section{Experimental Setup}

Previous investigations into the detection of ASR hallucinations have often relied on non-speech audio predictions \cite{baranski2025investigationwhisperasrhallucinations, calmwhisper}, synthetic noise injections \cite{baranski2025investigationwhisperasrhallucinations, frieske2024hallucinationsneuralautomaticspeech}  or proxy metrics \cite{fang2025fewerhallucinationsverificationthreestage, whisperx, tripathi2025listenliketeachermitigating, crisperwhisper} to judge effectiveness due to a lack of human-verified, real-speech hallucination data. In this study, we utilize the recently introduced HALAS
dataset \cite{baranski2026halas} which provides human-annotated hallucination labels for predictions generated by seven state-of-the-art ASR models using Earnings-22 audio data \cite{e22}. We focus on the Whisper large v3 model with 858 out of 3611 predictions marked as hallucinations (23.8\%), 18 marked as hallucinations containing looping, i.e. repeated phrases, (0.5\%), and the rest without hallucinations.

While HALAS provides span-level annotations, we focus on utterance level detection, and for all subsequent detector training and evaluation, we utilize the predefined train and test splits of this dataset. Moreover, for cross-validation grid search, the dataset is split into five folds, stratified by hallucination rate and audio duration, ensuring that recordings from the same speaker are grouped together. Experimental evaluation is performed using the Area Under the Receiver Operating Characteristic Curve (ROC AUC) together with standard classification metrics such as accuracy, precision, recall and F1 score.

\section{Proposed Detection Frameworks}

\subsection{Analysis based on text metrics}

First, we investigate the discriminative power of text-based metrics for hallucination estimation, categorizing them into oracle (reference-dependent) and reference-free methods.

\vspace{-3pt}
\subsubsection{Feature definitions}
\vspace{-3pt}

The most commonly utilized metrics require a ground-truth transcript to measure deviations in the ASR hypothesis. We utilize traditional error metrics, including Word Error Rate (WER), Character Error Rate (CER), Insertion Error Rate (IER) \cite{levenshtein1966binary}, and the reference vs. prediction Length Ratio (LenR) \cite{length_ratio}. To capture deeper semantic divergence,
we also employ BERTScore (BERT) \cite{bert} and SeMaScore (SeMa) \cite{Sasindran_2024}. Furthermore, we utilize the list of hallucinations from \cite{baranski2025investigationwhisperasrhallucinations}, which, through non-speech audio inference, established a model's specific hallucination
outputs. Based on this, we propose a Common Hallucinated Phrase (CHP) detector that flags an utterance if a known common pathological phrase appears in the prediction but is absent from the ground truth.

\begin{table}[!b]
\vspace{-12pt}
\caption{Hallu. detection ROC AUC for individual text features.}
\centering
\vspace{-10pt}
\resizebox{\columnwidth}{!}{%
\begin{tabular}{@{}l|cccccccc @{}}
\toprule
\textbf{Oracle} & BERT & CER & IER & WER & SeMa & LenR & CHP \\
\textbf{AUC [\%]} & 82.3 & 81.9 & 81.2 & 80.6 & 79.7 & 78.2 & 77.6 \\ 
\midrule
\textbf{Ref-free} & CPS & PPL & Align & NCHP & StopW & Rep & \multicolumn{1}{c}{} \\
\textbf{AUC [\%]} & 68.2 & 66.0 & 60.8 & 53.8 & 50.9 & 50.3 & \multicolumn{1}{c}{} \\
\bottomrule
\end{tabular}%
}
\label{tab:roc_auc_compact}
\vspace{-15pt} %
\end{table}

Reference-free metrics are less commonly used but are particularly important, as operating solely on the ASR prediction and audio metadata makes them applicable in zero-shot deployments. To identify degenerative text loops and linguistic anomalies, we extract the 5-gram Repetition rate (Rep) \cite{whisperx}, Stopword Ratio (StopW), and GPT 2 Perplexity (PPL) \cite{perplexity}. We also calculate Characters Per Second (CPS) to detect unnaturally dense text generation, and utilize wav2vec Forced Alignment (Align) \cite{fc} confidence scores to measure the temporal consistency between the generated text and the acoustic signal. Finally, we implement a Naive CHP detector (NCHP). Unlike its oracle counterpart, the naive approach simply checks for the presence of common erroneous phrases in the ASR prediction.

\vspace{-3pt}
\subsubsection{Capabilities of hallucination estimation}
\vspace{-3pt}

To evaluate the feasibility of lightweight hallucination detection, we begin with analysis of the discriminative power of individual text features. Table \ref{tab:roc_auc_compact} presents their ROC AUC scores in a hallucination detection task on the HALAS dataset.

Individual oracle metrics show strong discriminative capacity across semantic (BERT, AUC: 82.3\%) and structural (CER, AUC: 81.9\%) dimensions, alongside the CHP detector (AUC: 77.6\%). In contrast, no single reference-free feature is highly predictive on its own. CPS provides the strongest zero-shot performance (AUC: 68.2\%), capturing the unnaturally dense text generation typical of hallucinations and looping, closely followed by language model Perplexity (AUC: 66.0\%). To overcome single metric thresholding limitations, we propose fusing these varied features into a lightweight classification model.

\vspace{-3pt}
\subsubsection{Hallucination detection using text features}
\vspace{-3pt}

\begin{figure}[!t]
\begin{minipage}[b]{1.0\linewidth}
  \centering
  \centerline{\includegraphics[width=8.5cm]{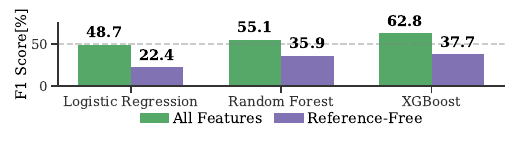}}
\end{minipage}
\vspace{-28pt}

\caption{F1 scores for text metric hallucination detection across classifiers using all vs reference-free features.}
\label{fig:text_detection_f1}
\vspace{-15pt}
\end{figure}

To synthesize text features into a detection mechanism, we trained three machine learning classifiers: Logistic Regression \cite{lr}, Random Forest \cite{randomforest}, and XGBoost \cite{xgboost}. We used the HALAS training split and optimized the models via Recursive Feature Elimination (RFE) \cite{rfe} for a setting that utilized all available feature values and a strictly reference-free setting for zero-shot deployment. Results are shown in Fig. \ref{fig:text_detection_f1}. Both tree-based models outperformed the Logistic Regression baseline (F1: 48.7\%), with XGBoost achieving the highest overall performance (F1: 62.8\%, Recall: 74.1\%), confirming that hallucination detection benefits from relying on complex, non-linear interactions between structural and semantic features.

Restricting the models to reference-free features resulted in severe performance degradation, with the top-performing XGBoost model dropping to an F1 score of 37.7\%. Most notably, without semantic and structural reference anchors, the models became overly conservative, suffering a collapse in Recall (Logistic Regression Recall: 15.5\%). While reference-free text features provide a computationally trivial mechanism for filtering the most egregious predictions, these results demonstrate that text-based hallucination detection currently remains highly dependent on ground-truth text.

\subsection{Hallucination detection using LLMs}
\label{llm_sec}
\begin{figure}[!t]
\vspace{-20pt}
\begin{minipage}[b]{1.0\linewidth}
  \centering
  \centerline{\includegraphics[width=8.5cm]{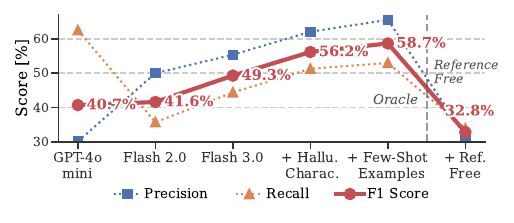}}
\end{minipage}
\vspace{-25pt}

\caption{Impact of iterative prompt enhancements and reference availability on LLM hallucination detection performance.}
\label{fig:llm_detection_f1}
\vspace{-20pt}
\end{figure}

We reproduced the approach from \cite{atwany_lost_2025} as a zero-shot baseline, asking the model to strictly isolate semantic fabrications from acoustic mishearings using the prompt provided by the authors. It explains the difference between hallucination and non-hallucination errors, provides five examples and asks the model to return a binary detection label. We evaluated GPT-4o mini\footnote{\url{https://developers.openai.com/api/docs/models/gpt-4o-mini}} and Gemini 2.0 Flash\footnote{\url{https://docs.cloud.google.com/gemini-enterprise-agent-platform/models/gemini/2-0-flash}}, and then iteratively enhanced the method in three steps. First, we increased the model's reasoning capacity by upgrading to Gemini 3.0 Flash\footnote{\url{https://docs.cloud.google.com/gemini-enterprise-agent-platform/models/gemini/3-flash}}. Second, we injected domain-specific pathology data by adding Whisper large v3 non-speech audio list of hallucinations \cite{baranski2025investigationwhisperasrhallucinations} as an error characteristic. Next, we included 10 targeted few-shot examples from the HALAS train split. Finally, we adapted the prompt for a strictly reference-free setting by removing the ground-truth text and shifting the detection criteria towards intrinsic textual anomalies and error pattern recognition. Figure \ref{fig:llm_detection_f1} presents the detection results across all configurations\footnote{LLM prompts for all configurations are available at \url{https://github.com/DSP-AGH/asr_hallucination_detection_prompts}}.

Out-of-the-box LLMs struggle significantly, with both achieving an F1 score of roughly 41\%. Flash 2 achieved a low Recall rate of 35.7\%, while GPT-4o-mini identified more hallucinations, with a higher Recall of 62.6\%. However, it suffers from an increased false positive rate, with poor precision 30.1\%. Upgrading to a stronger reasoning model (Flash 3.0) yielded a notable improvement, raising the F1 score to 49.3\%. More substantial gains were achieved through domain-specific prompt modifications: injecting known Whisper hallucination characteristics increased the F1 score by 6.9 percentage points (56.2\%). Combining this failure taxonomy with domain-specific few-shot examples yielded the highest overall LLM performance (F1: 58.7\%, Accuracy: 88.4\%).

Despite extensive prompt engineering and in-context learning, LLMs failed to surpass the XGBoost text-feature classifier (F1: 62.8\%). Given that both approaches require access to the ground-truth reference text, utilizing an LLM incurs massive computational overhead and latency without yielding a performance advantage. Finally, when transitioning to the reference-free setting, LLM performance suffered a collapse, dropping to an F1 score of 32.8\%. This mirrors the severe degradation observed in the reference-free text classifier, demonstrating that despite the extensive semantic knowledge embedded in LLMs, zero-shot hallucination detection remains fundamentally constrained by the absence of ground-truth anchors.

\subsection{Probing Whisper's decoder internal states}

Whisper utilizes a dynamic fallback decoding strategy. It begins with a 5-element beam search, but shifts to stochastic sampling with increasing temperature if the generation is deemed unconfident. This confidence is internally ranked using three heuristics: Average Log-probability (AL), Compression Ratio (CR), and No-Speech Probability (NSP). In theory, these values should flag unconfident, repetitive, or non-speech generations, respectively. We trained Logistic Regression classifiers to detect hallucination on these features, which proved highly ineffective, with an F1 score of 23.6\%. This shows that the values provided by Whisper fail at distinguishing hallucinations from general acoustic degradation. Thus, we investigated the embeddings generated during the decoding process. Previous work demonstrated that the final embedding, which contains the End-of-Sequence (EOS) token captures hallucination-related uncertainty \cite{glazer2025transcriptionmechanisticinterpretabilityasr}. We extended this approach by probing the sequence of embeddings at each decoder layer during inference.

\vspace{-3pt}
\subsubsection{Layer by layer hallucination information separability}
\vspace{-3pt}

\begin{figure}[!t]
\vspace{-18pt}
\begin{minipage}[b]{1.0\linewidth}
  \centering
  \centerline{\includegraphics[width=8.5cm]{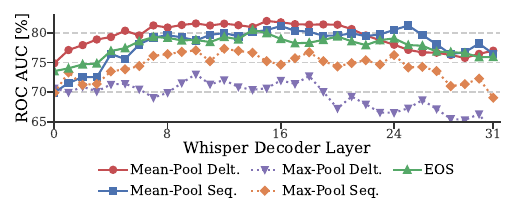}}
\end{minipage}
\vspace{-24pt}

\caption{Detection of hallucinations using 5-fold CV based on internal state probes (standard deviation $\leq$ 3 AUC points).}
\label{fig:roc_auc}
\vspace{-20pt}
\end{figure}

 To quantify the presence of hallucination information across Whisper’s decoder layers we employed linear probing via Logistic Regression and evaluated linear separability using ROC AUC. In addition to the final EOS embedding we analyzed the entire decoding sequence. To map embedding sequences, which are dependent on input length, to a fixed-size utterance vector, we conducted mean and max-pooling across decoding steps for both the sequence values and their deltas (step-by-step differences). Figure \ref{fig:roc_auc} plots the ROC AUC performance of these probes across all 32 decoder layers. Mean-pool deltas reached the highest AUC (82\%) at layer 15, with mean-pool embeddings achieving 81\% at layer 16 and EOS embedding 81\% at layer 14. This minimal variance between sequences, EOS and deltas suggests that hallucination markers are not localized in any single representation type. Max-pooling methods underperformed, demonstrating that hallucinations do not manifest as isolated spikes in embedding values but as systemic shifts.

Beyond the aggregation method, Fig. \ref{fig:roc_auc} reveals a distinct trajectory across the model's depth followed by all variants. Early layers (0-5) as well as late ones (27-31) perform significantly worse in separability, while middle layers show little variance. This indicates that hallucination information does not accumulate in a single layer or section of the decoder.

\vspace{-3pt}
\subsubsection{Hallucination detection using internal states}
\vspace{-3pt}

Building upon the promising separability of hallucinations in EOS embeddings (presented in Fig. \ref{fig:roc_auc}), we hypothesized that explicitly analyzing the entire decoding sequence could further improve detection. To capture these sequential dependencies, we utilized a Bidirectional Long Short-Term Memory (BLSTM) classifier on the entire decoding sequence embeddings \cite{blstm}. Whisper utilizes an encoder-decoder attention architecture, which at each layer allows for the extraction of projected embedding values from Self-Attention (SA), Cross-Attention (CA), and the final output after the Multi-Layer Perceptron (D). We compared the hallucination detection capabilities of embedding sequences extracted at these three points within the decoder block. For all evaluated internal-state approaches, optimal hyperparameters and up to two extraction layers were determined via Grid Search utilizing 5-fold cross-validation (CV). Furthermore, to model embedding transitions, we augmented the highest-performing representation by appending its sequential derivatives (deltas). Table \ref{tab:blstm} shows the best detector for each setting.

\begin{table}[!t]
\caption{Results of BLSTM classifiers based on 5-fold CV Grid Search (standard deviation between folds not exceeding 3.5\%).}
\centering
\vspace{-10pt}
\begin{tabular}{@{}l|ccccc@{}}
\toprule
\textbf{Features} & \textbf{D} & \textbf{SA} & \textbf{CA} & \textbf{$\Delta$ D} & \textbf{D+$\Delta$D} \\\midrule
\textbf{Layer num.} & (21,24) & (7,22) & (6,13) & (13,20) & (14,16) \\\midrule
\textbf{Acc. {[}\%{]}}  & 83.7 & 83.1 & 82.6 & 83.1 & 83.7 \\
\textbf{Prec. {[}\%{]}} & 67.9 & 65.8 & 65.7 & 67.6 & 68.1 \\
\textbf{Rec. {[}\%{]}}  & 64.8 & 63.9 & 62.7 & 60.6 & 63.4 \\
\textbf{F1. {[}\%{]}}   & 66.1 & 64.8 & 63.6 & 63.3 & 65.5 \\
\textbf{AUC. {[}\%{]}}   & 87.0 & 87.5 & 86.1 & 86.3 & 87.6 \\
\bottomrule
\end{tabular}
\label{tab:blstm}
\vspace{-15pt}
\end{table}

Results indicate that the middle-to-late layers of the Whisper decoder consistently provide the most discriminative signals for detecting hallucinations. The final output marginally outperformed both self-attention and cross-attention representations. Sequence deltas underperformed standard embeddings, suggesting they lack complementary predictive information.

The optimal BLSTM configuration, benchmarked using HALAS (i.e., trained using the train split and evaluated on the test), achieved an AUC of 87.6\% (F1: 65.5\%, Precision: 68.1\%, Recall: 63.4\%). This reference-free internal state detector outperforms the best text-feature models (F1: 62.8\%) despite not relying on ground-truth transcripts. By bypassing both the zero-shot performance collapse of text metrics and the computational overhead of LLMs, BLSTM classification on decoder internal states provides a more usable solution for real-world Whisper ASR hallucination detection.

\section{Detector Fusion}

\subsection{Comparison of different detection paradigms}

\begin{figure}[!hb]
\vspace{-8pt}
\begin{minipage}[b]{1.0\linewidth}
  \centering
  \centerline{\includegraphics[width=8.8cm]{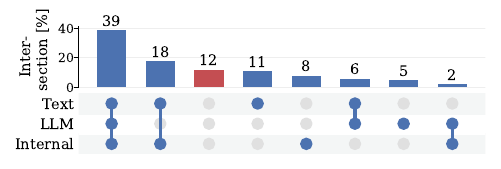}}
\end{minipage}
\vspace{-22pt}

\caption{Hallucination detections by detector combination. Dots show specific model groupings with corresponding bars indicating hallucinations detected exclusively by that combination. No dots highlight hallucinations missed by all models.}
\label{fig:overlap}
\vspace{-10pt}
\end{figure}

We analyzed detection overlap using pairwise agreement, revealing that the paradigms capture non-overlapping signals (agreement: 0.64-0.73).
The text classifier maximizes Recall (0.73) at the cost of Precision (0.53), while the LLM is the most conservative, with the highest Precision (0.64) but the lowest Recall (0.51). The internal state detector achieves the highest F1 score (0.62), being the most balanced. Figure \ref{fig:overlap} presents a correct detection comparison. All three models caught 39\% of examples, which were easier to detect multi-word fabrications. The 12\% missed by all three approaches were almost exclusively single-word function-word insertions (e.g., "was", "The"), which are indistinguishable without the audio signal.  

Hallucination phrase length is the dominant explanatory factor for performance. All models degrade on single-word hallucinations, but the LLM collapses most severely (Recall 0.40 vs. 0.67/0.59 for text/internal). Audio duration dependency further differentiates the paradigms. The internal classifier is the only one whose Recall improves to 0.75 on audio over 8 s, consistent with its ability to integrate evidence across a longer decoding sequence. The text detector performance peaks on files between 3-8 seconds, with a Recall of 0.80, indicating difficulty with longer files. This highlights an advantage of internal state detection in real-world long-form audio hallucination detection.

\subsection{Lightweight late fusion meta-classifier}

Our detector error analysis revealed that the XGBoost text-feature classifier and the BLSTM internal-state detector exhibit partially complementary failure modes. Due to weaker performance and computational overhead LLM detection is not included in this meta-classifier. We construct a late-fusion stacking ensemble that combines the outputs of both detectors into a lightweight meta-classifier. To prevent data leakage, we generate out-of-fold (OOF) predictions for the training set via 5-fold cross-validation. The meta-classifier receives three features: the hallucination probability from XGBoost, the hallucination probability from the BLSTM, and the audio duration in seconds. 
Results are presented in Table \ref{tab:fusion}.

\begin{table}[!t]
\caption{Late fusion results on the HALAS test split.}
\centering
\vspace{-10pt}
\begin{tabular}{@{}l|ccc@{}}
\toprule
\textbf{Model} & \textbf{Text} & \textbf{Internal} & \textbf{LR Meta} \\\midrule
\textbf{Acc. {[}\%{]}}  & 84.4 & 87.7 & 90.7 \\
\textbf{Prec. {[}\%{]}} & 54.4 & 58.1 & 71.0 \\
\textbf{Rec. {[}\%{]}}  & 74.1 & 66.7 & 65.7 \\
\textbf{F1. {[}\%{]}}   & 62.8 & 62.1 & 68.3 \\
\textbf{AUC. {[}\%{]}}  & 87.8 & 85.0 & 90.0 \\
\bottomrule
\end{tabular}
\label{tab:fusion}
\vspace{-15pt}
\end{table}

The Logistic Regression meta-classifier achieves the highest F1 score of 68.3\%, with a ROC AUC of 0.90, providing a sizable gain over single paradigm detectors. More complex meta-architectures (Decision Tree, Random Forest) did not improve F1. Notably, the analysis of the learned Decision Tree revealed that audio duration primarily appears at leaf-level boundary cases, contradicting our initial hypothesis that duration would serve as a primary routing signal between the two detectors. Ultimately, this late-fusion approach establishes the highest overall performance for hallucination detection on this dataset, highlighting the potential of cross-paradigm detection. While providing an improvement on internal-state detection, it does come at the cost of forfeiting reference-free usability.

\section{Conclusions}

Robust detection of ASR hallucinations remains challenging, particularly in zero-shot deployments where ground-truth references are unavailable. Our investigation across three paradigms reveals that while text-based classifiers and LLMs achieve strong oracle performance, both suffer performance collapses in strictly reference-free settings. We demonstrate that probing Whisper’s intermediate decoder states yields highly competitive detection via a BLSTM (F1 score of 62.1\%) without reference transcripts. While a late-fusion meta-classifier combining text and internal state probabilities achieves our highest overall performance (F1 score of 68.3\%) , it forfeits zero-shot usability. Ultimately, internal state probing, despite requiring direct model access, provides a highly promising solution. Our internal-state analysis is limited to Whisper, but this approach applies to other autoregressive encoder-decoder ASR models and, more broadly, to any architecture that exposes hidden states during token-by-token decoding. In contrast, the behavior of text-based and LLM-based detectors is model architecture agnostic. Extending internal-state probing to alternative ASR models and paradigms remains an important direction for future work. Furthermore, utilizing a specialized, lightweight LLM fine-tuned specifically for hallucination detection would bridge the gap between the zero-shot commercial LLMs tested and our supervised text-feature classifiers.

\section{Acknowledgments}

This research was supported by the National Science Centre, Poland under Grants 2021/42/E/ST7/00452 and 2023/49/B/ST7/04100, and by program "Excellence initiative – research university" for the AGH University of Krakow. We gratefully acknowledge Polish high-performance computing infrastructure PLGrid (HPC Centers: ACK Cyfronet AGH) for providing computer facilities and support within computational Grants PLG/2025/018238 and PLG/2025/018153. %

\section{Generative AI Use Disclosure}

The authors used large language models (ChatGPT, Gemini, Claude) to assist with language editing. All experimental design, data processing, statistical analysis, and scientific conclusions were independently conducted and verified by the authors. The authors take full responsibility for the content of this manuscript.

\bibliographystyle{IEEEtran}
\bibliography{mybib}

\end{document}